\documentclass[12pt,letter]{article}
\pdfoutput=1
\usepackage{graphicx, epsfig, color,cite}
\usepackage{amsmath}
\usepackage{amssymb}
\usepackage{float}
\usepackage{subfig}
\usepackage{hyperref}

\def\dslash{\not\!{\partial}}
\def\to{\rightarrow}

\def\bi{\begin{itemize}}
\def\ei{\end{itemize}}

\def\tchi{\tilde\chi}

\def\ta{\tilde a}

\def\tG{\tilde G}

\def\alt{\lesssim}
\def\agt{\gtrsim}
\def\be{\begin{equation}}  
\def\ee{\end{equation}}  
\def\bea{\begin{eqnarray}}  
\def\eea{\end{eqnarray}}

\begin{document}
\begin{titlepage}
\begin{flushright}
OU-HEP-260401
\end{flushright}

\vspace{0.5cm}
\begin{center}
      {\Large \bf Natural SUSY with mixed axion/axino dark matter}\\
\vspace{1.2cm} \renewcommand{\thefootnote}{\fnsymbol{footnote}}
{\large Howard Baer$^{1}$\footnote[1]{Email: baer@ou.edu },
  Vernon Barger$^2$\footnote[2]{Email: barger@pheno.wisc.edu},
  and Kairui Zhang$^1$\footnote[4]{Email: kzhang25@ou.edu}
}\\ 
\vspace{1.2cm} \renewcommand{\thefootnote}{\arabic{footnote}}
{\it 
$^1$Homer L. Dodge Department of Physics and Astronomy,
University of Oklahoma, Norman, OK 73019, USA \\[3pt]
}
{\it 
$^2$Department of Physics,
University of Wisconsin, Madison, WI 53706 USA \\[3pt]
}
\end{center}

\vspace{0.5cm}
\begin{abstract}
\noindent
While supersymmetric models provide a solution to the big hierarchy problem, natural SUSY is also allowed by the little hierarchy problem. 
In supersymmetric models which include the Peccei-Quinn (PQ) solution to the strong CP problem, one expects the presence of an axion-axino-saxion supermultiplet with a $\mu$eV-scale axion and a saxion with mass of order the soft breaking scale. The axino mass is much more model-dependent, and may occur in the range of keV-TeV: over 9 orders of magnitude. 
This leads to the possibility of the axino as lightest SUSY particle 
(LSP) and the presence of mixed axion plus axino dark matter. 
The case of natural SUSY with higgsino-like WIMPs as LSP seems (nearly) 
excluded by multi-ton noble liquid WIMP detector limits, even in the case
where the LSP has a depleted abundance compared to axions. 
We examine the case where the axino is LSP leading to mixed axion-axino 
dark matter in a natural SUSY context. 
We map out regions of PQ scale $f_a$ vs. axino mass $m_{\ta}$ parameter space where such a scenario remains viable in both the SUSY DFSZ and KSVZ axion models. For axino mass $\sim 100$ keV, we find 
solutions in accord with the measured dark matter abundance with mainly 
warm axino dark matter for $f_a\sim 10^{11}$ GeV and also solutions with
mainly axion cold DM and a tiny axino contribution for higher $f_a\sim 3\times 10^{12}$ GeV.

\end{abstract}
\end{titlepage}

\section{Introduction}

The Standard Model is beset by several finetuning problems: the gauge 
hierarchy problem (GHP: why is the weak scale so much smaller than the Planck or GUT scale), the strong CP problem (why is the CP-violating QCD Lagrangian contribution to the neutron electric dipole moment (EDM) 
so tiny) and the
cosmological constant problem (why is the spacetime vacuum energy
density over 120 orders of magnitude smaller than $M_{Planck}^4$?).
While the latter might be solved in the context of anthropic selection in an eternally inflating multiverse\cite{Weinberg:1988cp,Polchinski:2006gy}, the most promising solution to the first is softly broken supersymmetry\cite{Baer:2006rs} (SUSY) while the strong CP problem is solved by the Peccei-Quinn (PQ) mechanism and its concomitant 
{\it axion}\cite{Peccei:1977hh,Weinberg:1977ma,Wilczek:1977pj}.

While SUSY is adept at stabilizing the weak scale, thus solving the big hierarchy problem, the apparent lack of superpartners at the LHC (so far) has seemingly engendered a little hierarchy problem (LHP): 
why is there a mass gap between the weak scale and the superpartner mass scale? The LHP is quantified by large values of finetuning parameters 
$\Delta_i$, for $i=$ BG\cite{Barbieri:1987fn}, HS\cite{Kitano:2006gv} and EW\cite{Baer:2012up}. 
In Ref's \cite{Baer:2013gva,Baer:2023cvi,Mustafayev:2014lqa,Baer:2025uzx,Baer:2025zqt,Baer:2024fgd} it is argued that BG and HS overestimate finetuning
by factors of $10-1000$ as compared to the more conservative and model-independent measure $\Delta_{EW}$. Using $\Delta_{EW}$ with $m_h\simeq 125$ GeV, then it is found that many old favorite SUSY models such as CMSSM and GMSB
are indeed finetuned, but others such as gravity-mediation (as exemplified by non-universal Higgs models (NUHM)), natural 
anomaly-mediated SUSY breaking (nAMSB) and natural generalized 
mirage-mediation (nGMM) can have low values of $\Delta_{EW}\alt 30$, 
and so those portions of parameter space do not suffer from the LHP~\cite{Baer:2024tfo}, which has motivated dedicated collider studies of natural SUSY signatures guided by low values of $\Delta_{EW}$~\cite{Baer:2022qqr,Baer:2022smj,Baer:2023yxk,Baer:2023uwo,Baer:2023olq,Baer:2024hgq,Baer:2025uzx,Zhang:2026eoc,Baer:2025zqt}.

Of course, in pursuing plausible particle physics models beyond the SM, 
it is desirable to invoke solutions to {\it both} the GHP and the strong CP problem. Then one might envision the PQ-augmented Minimal Supersymmetric Standard Model (MSSM) as the correct low energy effective field theory below any high scales associated with further unifications.
One approach to solving the strong CP problem is to invoke intermediate
scale heavy quark fields which couple to PQ-charged objects, as in the 
so-called KSVZ models\cite{Kim:1979if,Shifman:1979if}. 
Another approach is to couple PQ charged gauge singlets to two Higgs doublets, the DFSZ approach\cite{Dine:1981rt,Zhitnitsky:1980tq}. 
Both KSVZ and DFSZ can be supersymmetrized, in which case the axion field $a$ is but one 
element of an axion superfield $A$ which contains in addition a 
spin-0 saxion $s$ and a spin-1/2 axino $\ta$. 

In addition to solving the strong CP problem, the axion provides an excellent candidate for cold dark matter (CDM) in our universe. 
In SUSY models with $R$-parity conservation (RPC)-- 
needed to stabilize the proton under dimension-4 operators--
the lightest SUSY particle (LSP), usually found to be the lightest neutralino $\tchi_1^0$, can also provide an excellent candidate for CDM
as a weakly interacting massive particle (WIMP).
Searches for WIMPs at multi-ton noble liquid detectors have recently placed formidable limits on WIMP dark matter\cite{LZ:2024zvo}, and these limits are
even approaching the so-called neutrino floor. Thus, many old favorite SUSY WIMP dark matter models such as well-tempered\cite{Arkani-Hamed:2006wnf} 
and focus-point\cite{Feng:2000gh} neutralinos are now ruled out\cite{Baer:2016ucr}. 
The light higgsinos of natural SUSY models, if they provide the entirety of dark matter, may also be ruled out\cite{Baer:2018rhs}.

By moving to the PQ-augmented MSSM (PQMSSM), one then gains
possibly two simultaneous DM candidates\cite{Baer:2011hx}: 
the lightest neutralino, a WIMP, and the axion $a$.
These mixed dark matter models have better accord with naturalness
since the thermally-produced light higgsino-like WIMPs typically make up only about 5-10\% of the DM abundance while axions make up the 
remainder\cite{Bae:2013bva}.
The lowered local abundance of WIMPs can bring the WIMP direct detection (DD) bounds back into accord with theory, but just barely\cite{Baer:2025zqt}.
Also, in the PQMSSM, the axion-photon-photon coupling is severely
suppressed by the presence of higgsinos in the axion anomaly couplings\cite{Bae:2017hlp}, so that SUSY axions lie well-below current axion haloscope search limits\cite{ADMX:2024xbv}.

In the present paper, we explore instead the possibility that the
axino $\ta$ is the LSP\cite{Rajagopal:1990yx}. 
Many early papers considered axino dark matter
in the case where axinos would make up the entirety of dark matter, and usually in the case where the underlying SUSY theory would now be considered as unnatural, such as in a CMSSM context\cite{Covi:1999ty,Covi:2001nw,Choi:2011yf}. 
In the CMSSM, where the lightest neutralino is usually bino-like, 
then WIMP dark matter is typically thermally overproduced\cite{Baer:2010wm}. 
However, with an axino as LSP, then the axinos may inherit the proto-WIMP abundance, as each WIMP could decay to
an axino, leading to\cite{Baer:2008eq,Baer:2009ms}
\be
\Omega_{\ta}^{NTP}h^2\simeq \frac{m_{\ta}}{m_{\tchi}}\Omega_{\tchi}h^2
\ee
thus bringing the presumed WIMP overabundance into accord with 
the measured DM abundance $\Omega_{DM}h^2\simeq 0.12$ for appropriate
values of the mass fraction $m_{\ta}/m_{\tchi}$. 
This is called  non-thermally-produced axinos (NTP). 

In a more realistic setting, axino dark matter should be accompanied also by cold axion dark matter produced by coherent axion field 
oscillations (CO)\cite{Abbott:1982af,Preskill:1982cy,Dine:1982ah}. 
Also, the axinos-- even though they are unlikely to be in thermal equilibrium due to their tiny coupling to matter suppressed by $1/f_a$-- can still be thermally produced (TP). 
The TP of axinos has been computed by Brandenberg and Steffen\cite{Brandenburg:2004du} 
and also by Strumia\cite{Strumia:2010aa} in the case of SUSY KSVZ where the axino couples to gluons via a derivative coupling which leads to a
linear dependence on the re-heat temperature $T_R$ arising from inflaton decay.

The TP axino production was computed in SUSY DFSZ in \cite{Bae:2011iw}, 
where the relic density is independent of $T_R$ due to the direct coupling of axinos to Higgs/higgsino fields. 

One must also account for TP and CO-produced saxions\cite{Bae:2013qr}. 
Saxions produced in the early universe can decay to WIMPs and also to axion pairs, thus increasing the DM abundance.
Even if these saxion decay modes to SUSY particles are suppressed, then saxion decays to SM particles inject entropy into the early universe which can dilute any DM relics present at the time of saxion decay.

We note that axino dark matter has also been considered recently with 
high scale SUSY\cite{Choi:2018lxt} and a KKLT setup\cite{Bae:2021rmg}; both these works ignore naturalness, unlike the present work.

\subsection{A natural SUSY benchmark point from the landscape}
\label{ssec:bm}

We adopt the  natural SUSY benchmark point NUHM3 as depicted in 
Table 1 of Ref. \cite{Baer:2024hpl}. 
In the string landscape, rather general arguments expect a power-law draw to large soft terms\cite{Douglas:2004qg,Susskind:2004uv,Arkani-Hamed:2005zuc} followed by anthropic selection of the 
scale for weak interactions in the ABDS window\cite{Agrawal:1997gf}: 
$0.5 m_{weak}^{OU}\alt m_{weak}^{PU}\alt 4m_{weak}^{OU} $ where $m_{weak}^{OU}\simeq m_{W,Z,h}\sim 100$ GeV is the weak scale in {\it our universe} (OU) and $m_{weak}^{PU}$ is the weak scale in each {\it pocket universe} (PU) within a ``friendly patch''\cite{Arkani-Hamed:2005zuc} of the greater multiverse.
The limits on the ABDS window, as required by the {\it atomic principle}
(that complexity arise in the forms of atoms as we know them) coincides 
with the naturalness measure\cite{Baer:2012up} $\Delta_{EW}\alt 30$\cite{Baer:2017uvn}.

The NUHM3 BM point arises from the four-extra-parameter non-universal Higgs model but where the (decoupled) first and second generation matter scalars are set equal to each other at $m_0(1,2)=30$ TeV. Third generation scalar soft masses are set at $m_0(3)=6$ TeV and unified
gaugino masses set to $m_{1/2}=2.2$ TeV. A large, negative $A_0$ term, which boosts $m_h\to 125$ GeV is given as $A_0=-6$ TeV and the
ratio of Higgs vevs $\tan\beta =10$. Finally, we set $m_A=2$ TeV and $\mu =200$ GeV. The spectra as generated by Isasugra\cite{Baer:1994nc} is given in Table 1 of Ref. \cite{Baer:2024hpl} and is not repeated here. 
The 30 TeV first/second generation matter scalars provide a decoupling solution to the SUSY flavor/CP problems\cite{Baer:2019zfl} while gluinos are at 5.2 TeV and the higgsino-like lightest neutralino has $m_{\tchi_1^0}\sim 200$ GeV. 
This leads to a would-be neutralino relic density of $\Omega_{\tchi}\sim 0.01$. The model is EW natural with $\Delta_{EW}=25$. 
Other landscape-selected natural SUSY spectra are qualitatively similar to our BM point and so we expect the analysis presented here to be rather general.

\section{Mixed axion/axino dark matter in the natural SUSY DFSZ model}

\subsection{SUSY DFSZ model}

The DFSZ axion model invokes {\it two} PQ charged Higgs doublets $H_{u,d}$ (a type-II Higgs doublet model) coupled to a PQ-charged gauge singlet via non-renormalizable couplings.
Such a construct fits well with the MSSM since both require
type-II two Higgs doublets. 
Elevating fields to superfields, then the essential SUSY DFSZ superpotential coupling is given as
\be
W_{DFSZ}=\lambda_\mu X^2 H_u H_d/m_P
\ee
where $m_P$ is the reduced Planck mass, $H_{u,d}$ are the Higgs superfields carrying $PQ(H_{u,d})= +1$ and $X$ is the singlet carrying $PQ(X)=-1$. Note that the usual superpotential $\mu$-term is forbidden by global $U(1)_{PQ}$ symmetry. 
{\it Better yet}: if the $\mu$ term is forbidden by some more
fundamental discrete (gauge or $R$) symmetry, then the MSSM develops
an accidental global $U(1)_{PQ}$ which can be used to solve the strong CP problem~\cite{Baer:2025oid,Baer:2025srs,Sheng:2026aro,Sheng:2025sou}.
One can add a second PQ charged gauge singlet field $Y$ to the superpotential in order to stabilize the scalar potential: 
{\it e.g.}  $W\ni f/m_P X^3Y$. 
Under SUSY breaking, the scalar components of $X$ and $Y$ develop
soft terms $\sim m_{soft}$ which results in the $X$ and $Y$ fields
developing PQ breaking vevs $v_X$ and $v_Y$ 
$\sim \sqrt{m_{soft}m_P}\sim f_a$ thus 
\bi
\item developing the needed $\mu\sim \lambda_\mu f_a^2/m_P$ and
\item spontaneously breaking the global PQ so as to generate a pseudo-Goldstone boson, the axion.
\item Along with the axion field needed for the strong CP solution, 
an $R$-parity even spin-0 saxion $s$ and an $R$-odd spin-1/2 axino $\ta$ arises (from combinations of the $X$ and $Y$ fields). 
The saxion develops a mass $m_s\sim m_{soft}$ while the axino mass $m_{\ta}$ is expected of order $m_{soft}$ in simple gravity-mediation models but is model-dependent\cite{Goto:1991gq,Chun:1992zk,Chun:1995hc,Kim:2012bb} and can also be much lighter: as low as the keV scale. In this latter case, 
the axino could be the lightest SUSY particle and a candidate component of dark matter in the universe!\cite{Rajagopal:1990yx}
\ei
While the case of mixed neutralino/axion dark matter in the SUSY DFSZ model has been explored in Ref. \cite{Bae:2013hma}, here we examine the corresponding case for mixed axino/axion dark matter. 
The case of axino dark matter in SUSY KSVZ has been explored in 
Ref's \cite{Covi:1999ty,Covi:2001nw,Choi:2011yf}, while mixed axion/axino dark matter in SUSY KSVZ has been explored in Ref. \cite{Baer:2009ms}:
for a review, see {\it e.g.} \cite{Baer:2014eja}. These latter works have all taken place in the rather implausible case of unnatural SUSY typically with a bino NLSP.
Here we address mixed axion/axino dark matter from SUSY DFSZ in natural SUSY with a higgsino-like NLSP.

\subsection{Higgsino-like neutralino decay to axinos in SUSY DFSZ}

The $\tchi_1^0\ta Z$ and $\tchi_1^0\ta h$ couplings (and others) 
have been worked out in Ref. \cite{Bae:2013hma} (see Appendix A). Here, we note that for a natural, higgsino-like NLSP with $m_{\tchi_1^0}\sim \mu\sim 100-350$ GeV, then the dominant decay modes are expected to be 
1. $\tchi_1^0\to \ta Z$ and 2. $\tchi_1^0\to \ta h$ assuming the 2-body decays are kinematically allowed. 
The axino-Higgs-higgsino couplings are expected to have the form
\be
{\cal L}_{DFSZ}\ni \int d^2\theta (1+B\theta^2) \mu \exp(c_H A/v_{PQ}) H_u H_d
\ee
where $1+B\theta^2$ is a SUSY breaking spurion field, $c_H$ is a model-dependent coupling of order unity, $A$ is the axion supermultiplet
and $v_{PQ}^2=\sum_i q_i^2v_i^2$ with $q_i$ as PQ charges of the
PQ breaking fields $i$ which obtain vevs $v_i$\cite{Bae:2013hma}. 
The axion decay constant is then given as $f_a=\sqrt{2}v_{PQ}$.
The approximate decay width for decay to light Higgs $h$ and $Z$ is 
of the form
\be
\Gamma (\tchi_1^0\to \ta h,\ \ta Z)\sim \frac{c_H^2}{32\pi}\frac{\mu^2}{v_{PQ}^2}m_{\ta} .
\ee

\begin{figure}[htb!]
\centering
    {\includegraphics[height=0.4\textheight]{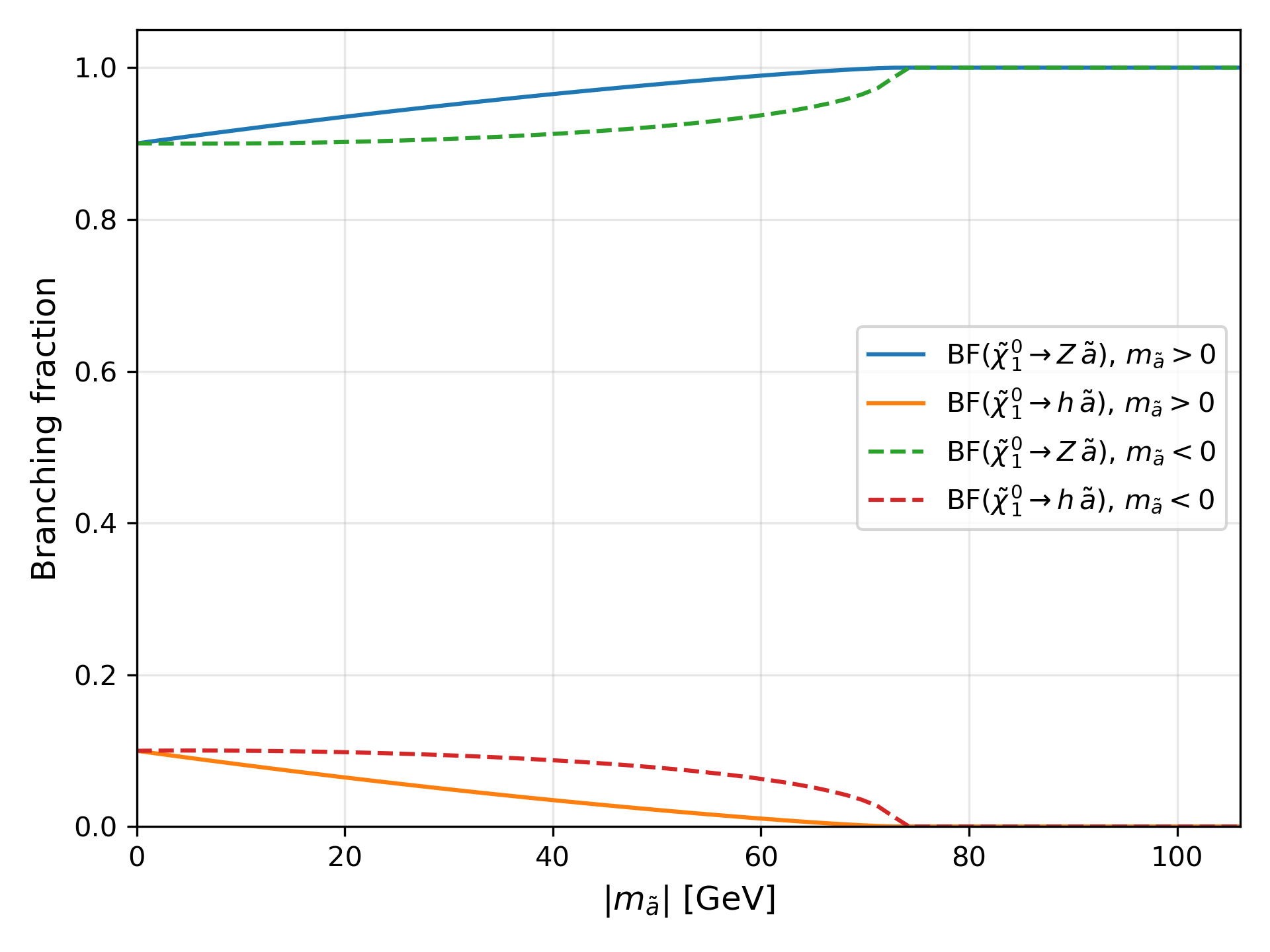}}
        \caption{Branching fraction of lightest neutralino $\tchi_1^0$
        into $Z\ta$ and $h\ta$ versus $m_{\ta}$ in the 
        SUSY DFSZ axion model for both positive and negative $m_{\ta}$.
      \label{fig:bfz1}}
\end{figure}

If two-body decays are closed, then the 3-body decays
via $Z^*$ and $h^*$ will occur. One can also have decays $\tchi_1^0\to \ta \gamma$ but these proceed via the mixing-suppressed bino components of $\tchi_1^0$. The exact tree-level decay formulae, using crossing symmetry, are extracted from Equations A.49 and A.75 of Ref. \cite{Bae:2013hma}. 
The branching fractions are displayed in Fig. \ref{fig:bfz1} vs. $m_{\ta}$ for the natural SUSY benchmark point BM1 from Subsec. \ref{ssec:bm} for positive and negative values of $m_{\ta}$.
Typically, $\tchi_1^0$ decay into real $Z$ bosons will dominate the decay into real Higgs $h$.

An important restriction on late-decaying neutral particles in the early universe comes from requiring their decay before the onset of Big-Bang nucleosynthesis (BBN). These constraints are mapped out in 
the putative relic density of the late decaying neutral particle
$\Omega_{\tchi}h^2$ vs. $\tau_{\tchi}$ plane for various assumed hadronic branching fractions of the decaying particle\cite{Jedamzik:2006xz}. 
From Fig. 10 of Ref. \cite{Jedamzik:2006xz}, with $\Omega^{TP}_{\tchi}\sim 0.01$, we would then expect $\tau(\tchi )\alt 10^2$ s.

\begin{figure}[htb!]
\centering
    {\includegraphics[height=0.4\textheight]{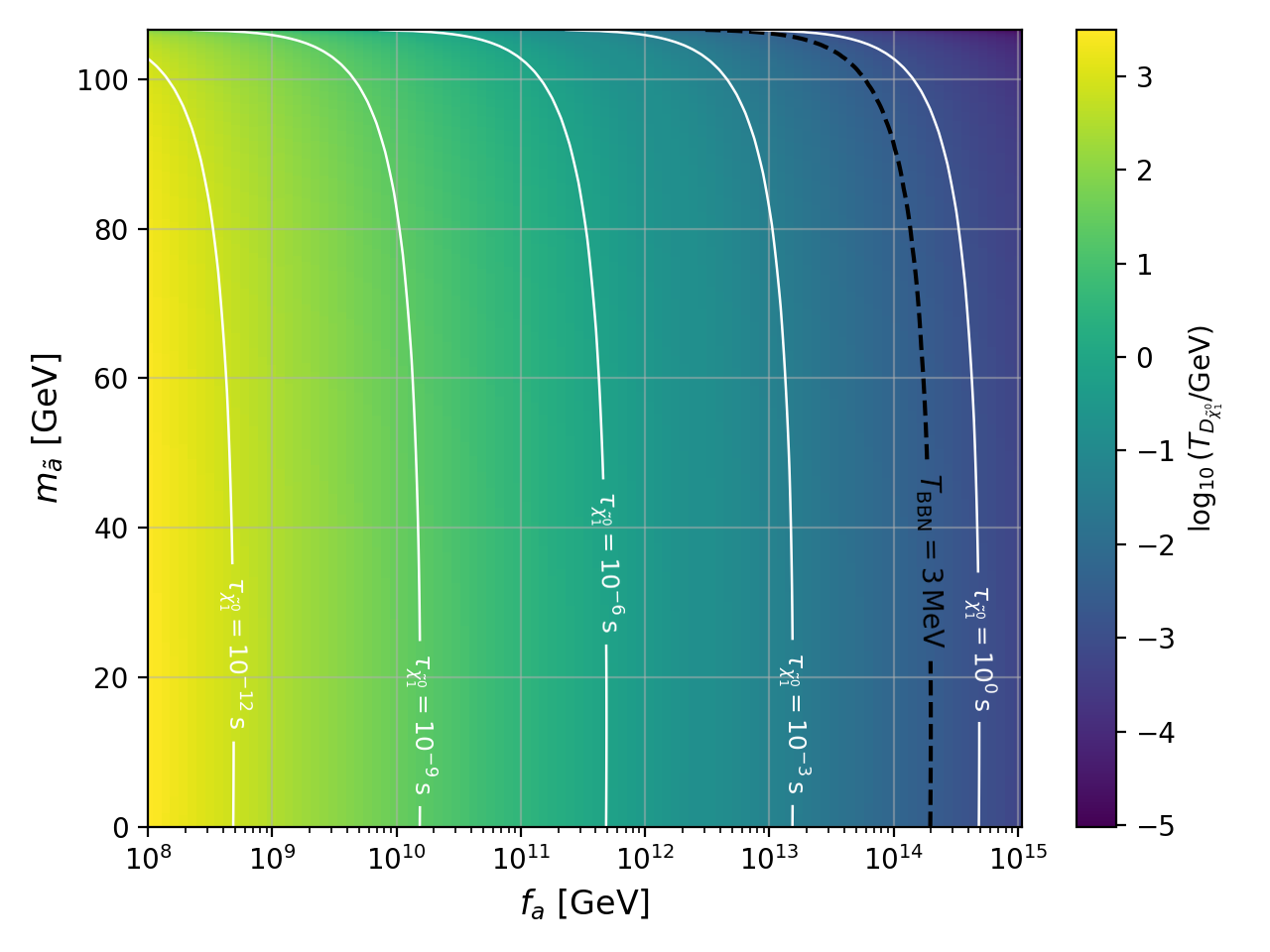}}
        \caption{Decay temperature $T_D(\tchi_1^0)$ in the 
        $f_a$ vs. $m_{\ta}$ plane. We also show the rough bound for the
        $\tchi_1^0$ to be BBN-safe, {\it i.e.} that it decays before the onset of BBN. In white are shown contours of $\tau(\tchi_1^0 )$ ranging from $10^{-12}--\ 1$ s.
      \label{fig:TDz1}}
\end{figure}

Here, we compute the decay temperature $T_D$ of the lightest neutralino
\be
T_D(\tchi_1^0 )=\frac{\sqrt{\Gamma_{\tchi} m_P}}{(\pi^2g_*(T_D)/90)^{1/4}}
\ee
and to simplify merely require that $T_D(\tchi_1^0 )\alt 3-5$ MeV.
The results are plotted in Fig. \ref{fig:TDz1} in the $m_{\ta}$ vs. $f_a$ plane
for the case of our BM point. The neutralino decays tend to occur in the MeV-GeV range and can become BBN-excluded for values of $f_a\agt 10^{15}$ GeV. In white are shown contours of neutralino lifetime $\tau(\tchi_1^0 )$ ranging from $10^{-12}-\ 1$ s.
Neutralinos with lifetimes within this range may be susceptible to special signatures endemic to long-lived particles (LLPs) 
at a bevy of LLP search experiments such as ATLAS/CMS, FASER and MATHUSLA\cite{Curtin:2018mvb,Alimena:2019zri,Barenboim:2014kka,Jeanty:2026etw}.

\section{Production of mixed axion/axino dark matter in the natural 
DFSZ SUSY model}
\label{sec:prod}

The production of mixed axion/axino dark matter in the SUSY DFSZ model
can be intricate since there are a variety of dark matter production processes available. 
\begin{enumerate}
\item First, there is the usual axion production via axion field coherent oscillations $\Omega_a^{CO} h^2$.
\item Next, there is non-thermal production (NTP) of axinos via thermal
neutralino production followed by neutralino decays to axinos. Here, the NTP axinos inherit the neutralino number density so that 
$\Omega_{\ta}^{NTP}h^2=\frac{m_{\ta}}{m_{\tchi}}\Omega_{\tchi}h^2$.
\item Using typical values of $f_a\sim 10^{11}$ GeV, the axinos are never in thermal equilibrium, and yet they can be produced thermally (TP) via the {\it freeze-in} mechanism\cite{Hall:2009bx}. 
The TP axinos start with minimal abundance at high temperatures, and then develop an abundance from sparticle decays to axinos and axino pair production in the early universe: $\Omega_{\ta}^{TP}h^2$.
\item One may also include TP gravitinos, where 
$\Omega_{\tG}^{TP}h^2 \sim T_R$, the reheat temperature after inflaton decay. The gravitinos cascade down to axinos, so that this population of axinos inherit the thermally-produced gravitino number density.
\item Lastly, it is possible to produce saxions either thermally or via COs. The saxions may decay to SM particles, to SUSY particles (which cascade down to the $\tchi$ state), to axions $s\to aa$ (resulting in dark radiation) or directly to axinos $s\to\ta\ta$.
\end{enumerate}
The axion-axino-saxion kinetic terms and self-couplings (in four component notation) are of the form
\be
{\cal L}\ni \left( 1+\frac{\sqrt{2}\xi}{v_{PQ}}s\right)
\left[\frac{1}{2}\partial^\mu a\partial_\mu a +\frac{1}{2}\partial^\mu s\partial_\mu s + \frac{i}{2}\bar{\ta}\dslash\ta\right]
\ee
where $\xi =\sum_i q_i^3v_i^2/v_{PQ}^2$. The $\xi$ value is model dependent and typically of order $\sim 1$ but can also be nearly zero\cite{Chun:1995hc}.

The total mixed axion/axino dark matter relic density is then given by
\be
\Omega_{a\ta} h^2=\Omega_a^{CO} h^2+\frac{m_{\ta}}{m_{\tchi_1^0}}\Omega_{\tchi_1^0}^{TP}h^2+\Omega_{\ta}^{TP}h^2+\frac{m_{\ta}}{m_{\tG}}\Omega_{\tG}^{TP}h^2 +\Omega_{a\ta}h^2(saxion).
\ee

The TP axino abundance for SUSY DFSZ has been calculated in {\it e.g.} \cite{Bae:2011jb,Chun:2011zd,Bae:2011iw}
and is given by
\be
\Omega_{\ta}^{TP}h^2\simeq 0.11\frac{m_{\ta}}{18\ {\rm keV}}\frac{\mu}{1\ {\rm TeV}}\left( \frac{10^{11}\ {\rm GeV}}{f_a/c_H} \right)^2 .
\ee
Here, the expression is independent of $T_R$, unlike the case of SUSY KSVZ. We immediately see from this expression that for $f_a\sim 10^{11}$ GeV, then to avoid overclosure, the axinos must be very light: of order
keV-MeV values for typical ranges of $f_a$. With such light axinos, 
then the NTP axinos, whose abundance is suppressed by the mass ratio
$m_{\ta}/m_{\tchi}$, have a negligible abundance. 

For the axion abundance, we use the standard axion abundance as expected from CO-production\cite{Visinelli:2009zm}:
\be
\Omega_a^{CO} h^2\simeq 0.23 f(\theta_i)\theta_i^2\left(\frac{f_A/N}{10^{12}\ {\rm GeV}}\right)^{7/6}
\ee
with $N=6$ for DFSZ and
\be
f(\theta_i )= \left[ \ln\left(\frac{e}{1-\theta_i^2/\pi^2}\right)\right]^{7/6}
\ee
is the anharmonicity factor.

We also include in our calculations the thermal production of gravitinos in the early universe. 
We here follow Pradler and Steffen, who  have estimated the thermal gravitino production abundance as~\cite{Pradler:2006qh}
\be
\Omega_{\tG}^{\rm TP}h^2 =\sum_{i=1}^{3}\omega_ig_i^2(T_R)
\left(1+\frac{M_i^2(T_R)}{3m_{\tG}^2}\right)\ln\left(\frac{k_i}{g_i(T_R)}\right)
\left(\frac{m_{\tG}}{100\ {\rm GeV}}\right)\left(\frac{T_R}{10^{10}\ {\rm GeV}}\right) ,
\ee
where $\omega_i=(0.018,0.044,0.117)$, $k_i=(1.266,1.312,1.271)$, $g_i$ are the gauge  couplings evaluated at $Q=T_R$ and $M_i$ are the gaugino masses also evaluated at $Q=T_R$. The axino abundance from gravitino 
decay is likewise suppressed by a factor $m_{\ta}/m_{3/2}$ so our plots
for SUSY DFSZ will have hardly any dependence on $T_R$.

We will assume $\xi =0$ for saxion couplings to axions/axinos since
no apparent excess of dark radiation is evident in the latest count of
$N_{eff}=2.99\pm 0.17$\cite{ParticleDataGroup:2024cfk}, whereas the SM value is $N_{eff}=3.045$, 
leaving almost no room for extra dark radiation. Saxions can still decay to SUSY particles which cascade down to axinos. But if saxions decay before neutralino freeze-out, then this population is washed out.
For our SUSY BM point, we expect $m_s\sim 30$ TeV and so very large
values of $f_a$ are required for $s$ to decay after neutralino freeze-out. Here, we will ignore the contribution to the dark matter abundance from saxions. 
The saxion decay formulae, branching fractions and decay temperature $T_D(s)$ for SUSY DFSZ are shown in Ref. \cite{Bae:2013hma}. 
For a complete treatment of intertwined axion-axino-saxion-neutralino-gravitino effects on dark matter, an eight-coupled Boltzmann equation solution is needed, similar to Ref. \cite{Bae:2014rfa}.
Thus-- for our results here-- we set $\Omega_{a\ta}h^2(saxion)\to 0$.

\begin{figure}[htb!]
\centering
    {\includegraphics[height=0.4\textheight]{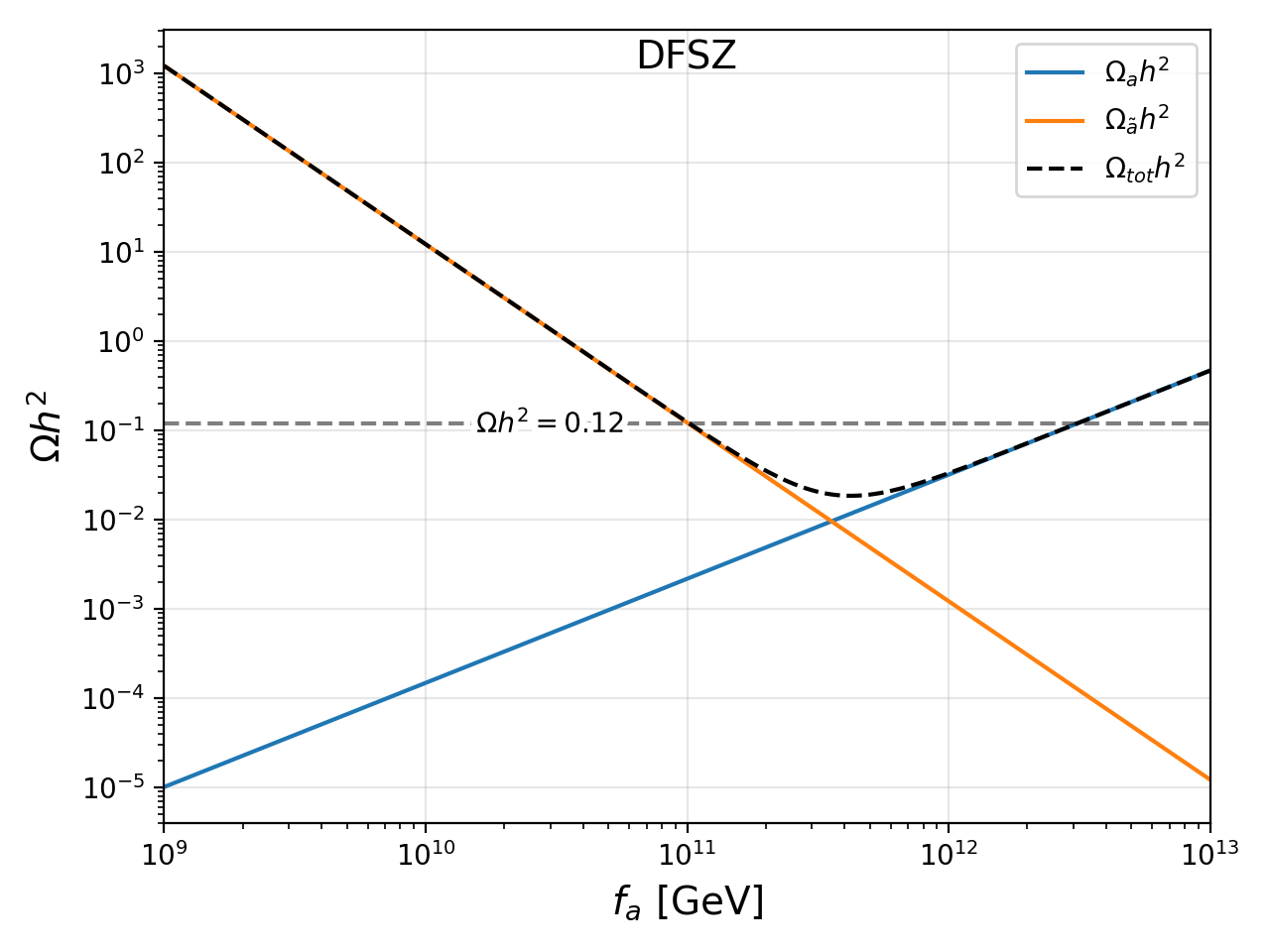}}
        \caption{Mixed axion-axino relic abundance vs. $f_a$ 
        for $m_{\ta}=100$ keV in the benchmark SUSY DFSZ model with $m_{3/2}=10$ TeV and $\theta_i=1$.
      \label{fig:Oh2_ata_DFSZ}}
\end{figure}

Our first results for the mixed axion/axino abundance are shown in Fig. \ref{fig:Oh2_ata_DFSZ} where we show $\Omega_{a\ta}h^2$ vs. $f_a$ for 
axino mass $m_{\ta}=100$ keV, $m_{3/2}=10$ TeV, $\theta_i=1$ and $T_R=10^6$ GeV. The plot is dominated by TP-axinos on the left and CO-produced axions on the right. The TP axinos have a tremendous abundance for smaller values of $f_a$ due to the increased axino coupling constant.
At two values A.) $f_a$-- $\sim 10^{11}$ GeV and B.) $\sim 3\times 10^{12}$ GeV-- the mixed axion/axino relic density is brought into accord with the measured value $\Omega_{DM}h^2=0.12$. For $f_a\simeq 10^{11}$ GeV, 
the dark matter abundance is axino-dominated with axions contributing at the 2-3\% level. At $f_a\sim 3\times 10^{12}$ GeV, then the DM abundance is axion-dominated with axinos contributing at $\sim 0.1\%$.
Since sub-GeV axinos (here with $m_{\ta}\sim $keV-MeV) are considered to be on the edge of warm DM candidates\cite{Jedamzik:2005sx,Bae:2017dpt}, we expect that solution B may be favored as it leads to 
cold DM axions as by far the bulk of dark matter, with only a tiny portion of axinos.
We note further that the axion abundance curve is proportional to $\theta_i^2$ so it can be dialed up or down from our assumed value of $\theta_i\sim 1$.

\begin{figure}[htb!]
\centering
    {\includegraphics[height=0.4\textheight]{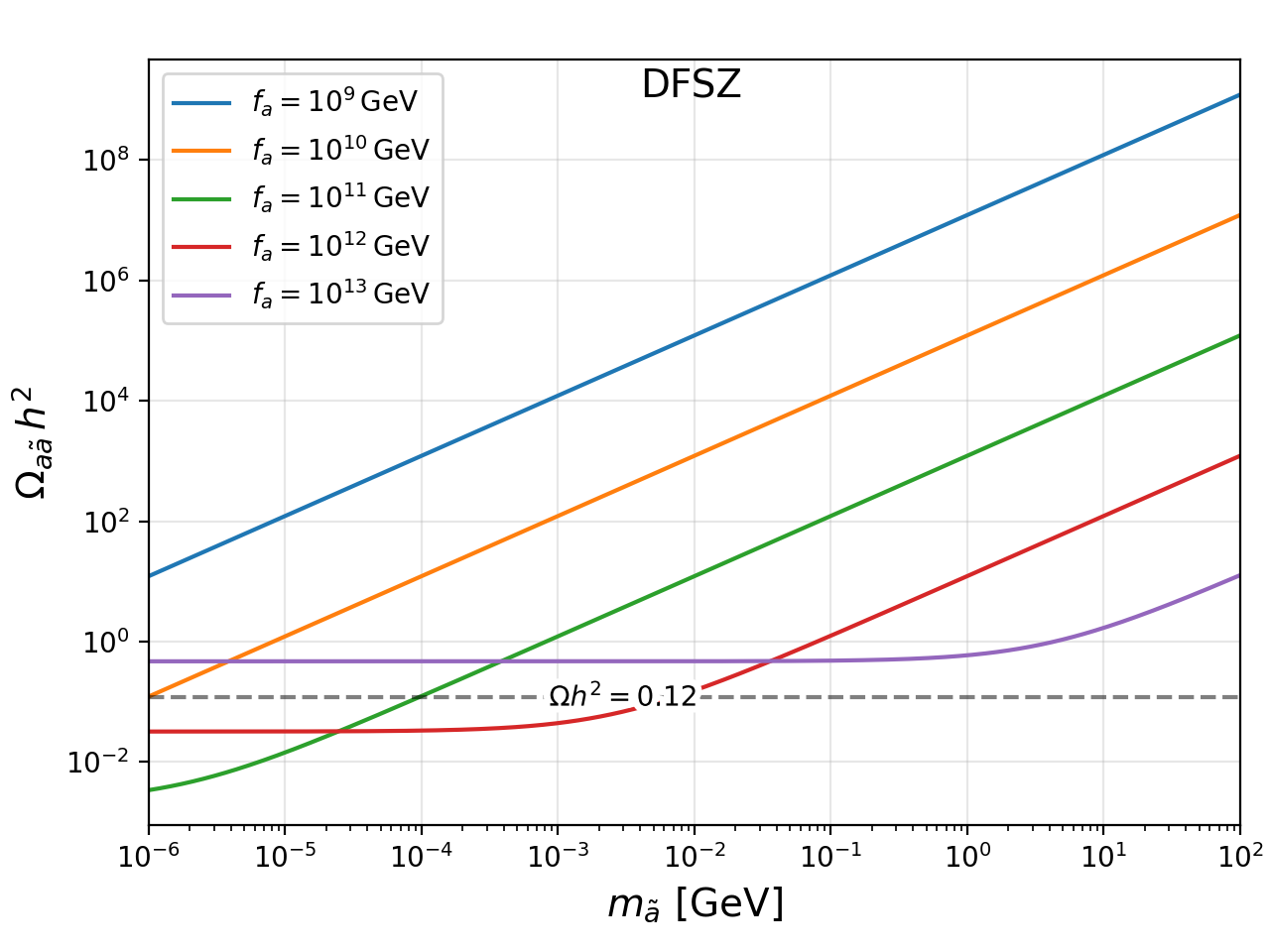}}
        \caption{Mixed axion-axino relic abundance vs. $m_{\ta}$ 
        for various values of PQ scale $f_a$ in the SUSY DFSZ benchmark model
        with $m_{3/2}=10$ TeV, $T_R=10^6$ GeV and $\theta_i=1$.
      \label{fig:Oh2_DFSZ}}
\end{figure}

Another view of the mixed $a\ta$ abundance is shown in Fig. \ref{fig:Oh2_DFSZ}
where we show again $\Omega_{a\ta}h^2$ but this time vs. $m_{\ta}$ for
five different $f_a$ values from $10^9-10^{13}$ GeV. The measured value 
is shown by the horizontal dashed line. For very small $f_a\sim 10^9$ GeV,  then axino DM is overproduced for all values of $m_{\ta}$ shown. Likewise, for $f_a\sim 10^{13}$ GeV (purple line), all values of
$m_{\ta}$ yield too much DM. The purple curve levels off at a constant value on the left side since that is where the axion abundance dominates, which is independent of $m_{\ta}$. 
The curve starts increasing at 
$m_{\ta}\agt 1$ GeV where the axino abundance starts becoming important. Of note here are solutions with $m_{\ta}\sim 10^{-4}$ GeV and 
$10^{-2}$ GeV: these two cases exhibit accord with the measured value $\Omega_{a\ta}h^2\sim 0.12$. 
The red curve with $f_a=10^{12}$ GeV and $m_{\ta}\sim 10^{-2}$ GeV is
axion-dominated for $m_{\ta}\alt 10^{-2}$ GeV while the green curve is axino-dominated for axino mass $\agt 10^{-5}$ GeV.

\begin{figure}[htb!]
\centering
    {\includegraphics[height=0.5\textheight]{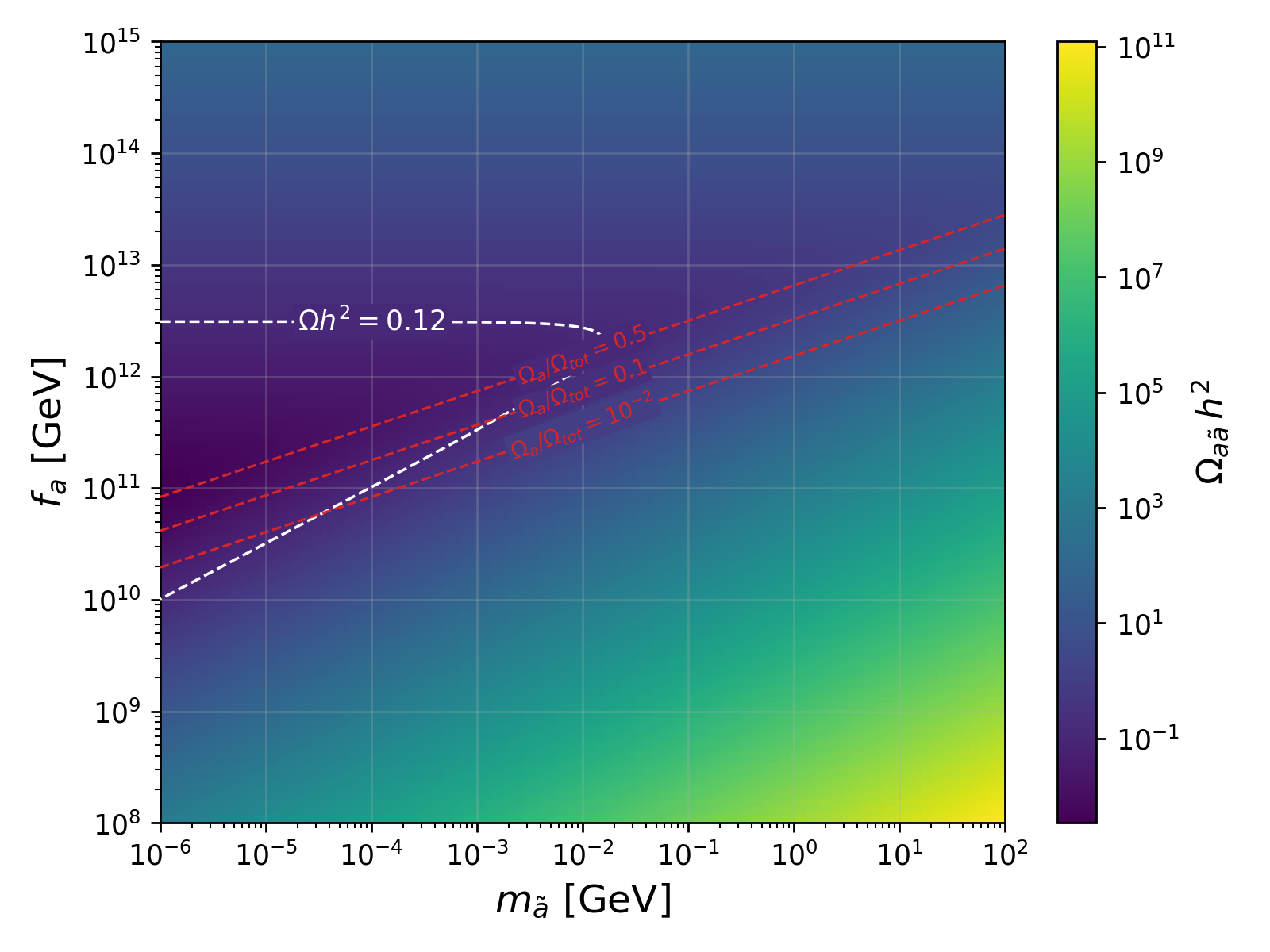}}
        \caption{Color-coded mixed axion-axino relic abundance 
        in the $m_{\ta}$ vs. $f_a$ plane 
        in the SUSY DFSZ model
        with $m_{3/2}=10$ TeV, $T_R=10^6$ GeV and $\theta_i=1$.
      \label{fig:Oh2_plane}}
\end{figure}

In Fig. \ref{fig:Oh2_plane}, we show the mixed axion/axino relic density in the $m_{\ta}$ vs. $f_a$ plane for $\theta_i=1$. 
The white contour shows where $\Omega_{a\ta}h^2\simeq 0.12$ and within
this contour one has $\Omega_{a\ta}h^2<0.12$, {\it i.e.} an underabundance  (although the axion portion can be easily increased via larger values of $\theta_i$). In the bulk of the plane, mixed axion/axino dark matter is overproduced (although the axion abundance can be dialed down with lower values of $\theta_i$). For $\theta_i=1$ and $m_{\ta}\agt 1$ MeV, then the mixed $a\ta$ dark matter is always over-abundant.
The red contours show ratios of the portion of axion abundance to the total abundance, and so the upper portion of the plane has mainly axion dark matter while the lower portion has mainly axino dark matter.

\section{Axino dark matter in natural SUSY with KSVZ axions}
\label{sec:ksvz}

\begin{figure}[htb!]
\centering
    {\includegraphics[height=0.4\textheight]{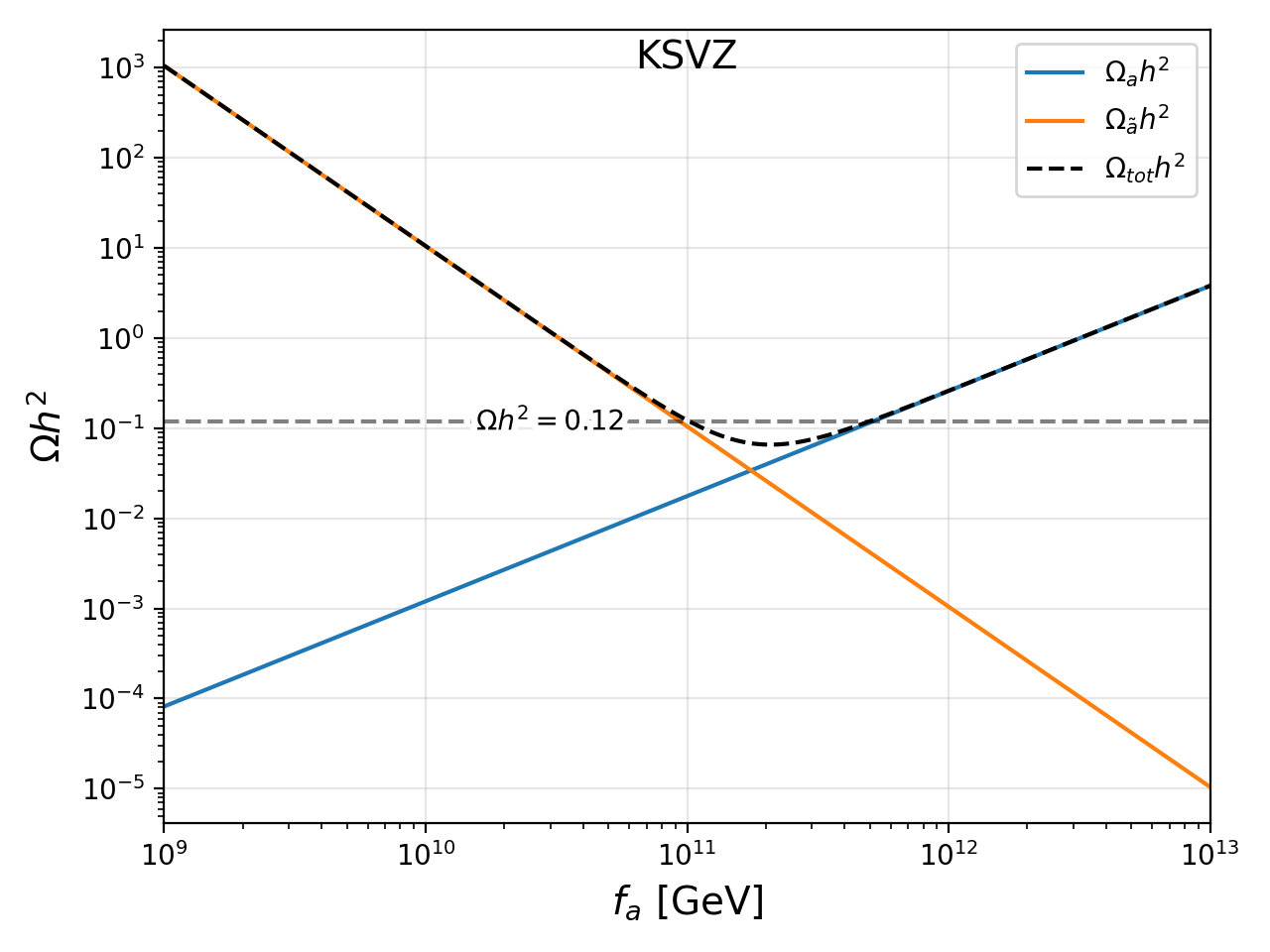}}
        \caption{Mixed axion-axino relic abundance vs. $f_a$ 
        for $m_{\ta}=100$ keV in the SUSY KSVZ model.
      \label{fig:Oh2_ata_KSVZ}}
\end{figure}

In the SUSY KSVZ model, the lightest neutralino decays via its bino component as $\tchi_1^0\to \ta\gamma$ with a rate that is governed by $(1/f_a)^2$ and with a lifetime of order $\tau_{\tchi}\sim 0.01-1$ sec\cite{Covi:1999ty}. The main difference from SUSY DFSZ is that the thermally-produced axino production rate depends linearly on $T_R$
due to the axino-gluino-gluon coupling and also that CO-produced
axions require $N_{DW}=1$ instead of 6.
We use the BS\cite{Brandenburg:2004du} calculation of the axino thermal production rate. 
Then the mixed axion/axino relic density $\Omega_{a\ta}^{KSVZ}h^2$ 
is given in Fig. \ref{fig:Oh2_ata_KSVZ} vs. $f_a$ for 
$m_{\ta}=100$ keV and $T_R=10^6$ GeV. 
The value of $\Omega_{\ta}^{TP}h^2$ can be dialed up or down depending on the assumed $T_R$ since its value depends linearly on $T_R$.
The plot is rather similar
to that of Fig. \ref{fig:Oh2_ata_DFSZ} with a mainly WDM axino solution at $f_a\sim 10^{11}$ GeV and a mainly axion CDM solution at 
$f_a\sim 5\times 10^{11}$ GeV. In the latter case, the WDM axinos only make up $\sim 3\%$ of the DM relic density.

\begin{figure}[htb!]
\centering
    {\includegraphics[height=0.4\textheight]{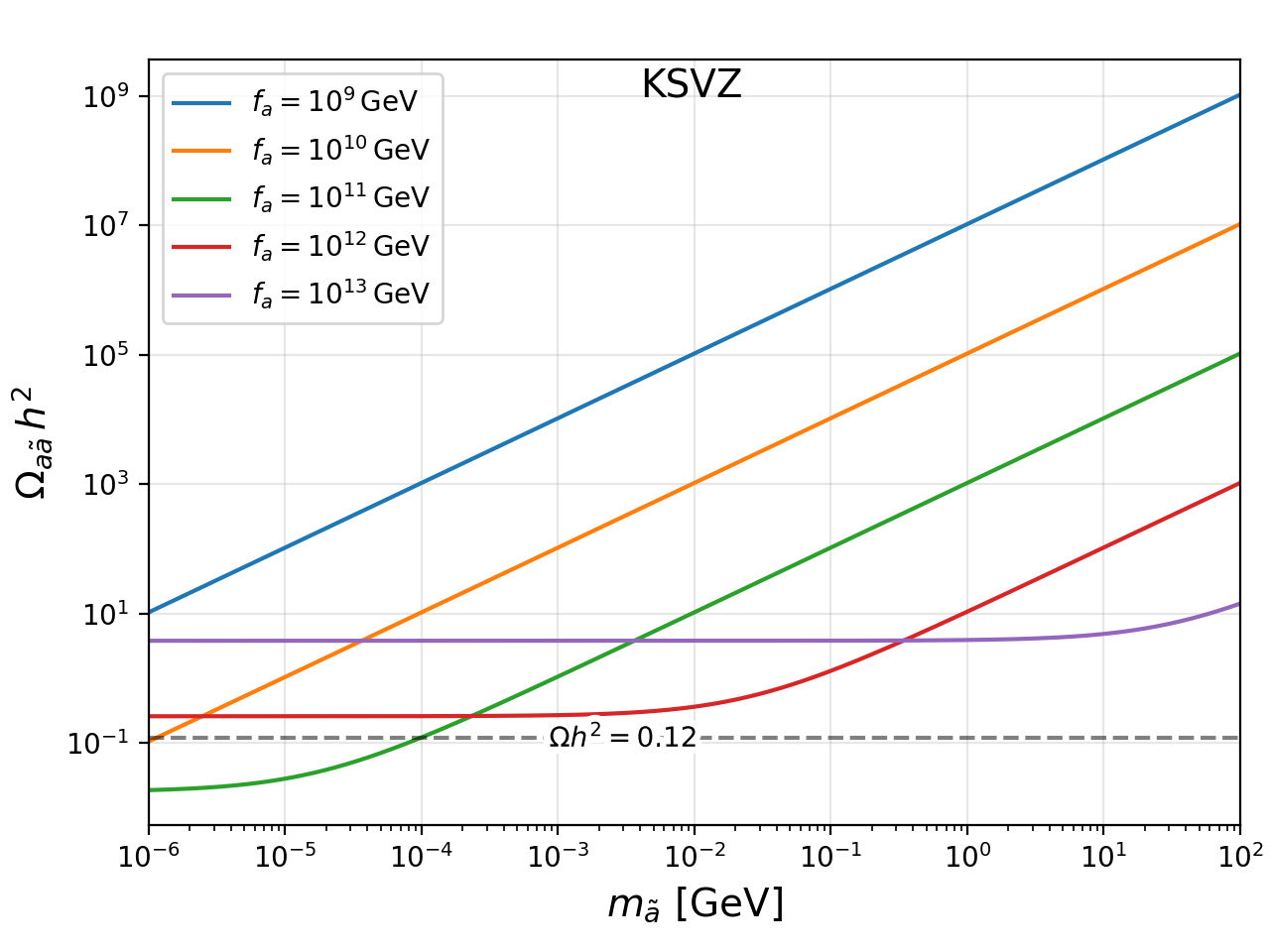}}
        \caption{Mixed axion-axino relic abundance vs. $m_{\ta}$ 
        for various values of PQ scale $f_a$ in the SUSY KSVZ model
        using thermal axino production ala Brandenberg \& Steffen,
        Ref. \cite{Brandenburg:2004du}.
      \label{fig:Oh2_Steffen}}
\end{figure}

In Fig. \ref{fig:Oh2_Steffen}, we show the value of $\Omega_{a\ta}^{KSVZ}h^2$ vs. $m_{\ta}$ for the same five values of $f_a$ as in Fig. \ref{fig:Oh2_DFSZ}. For low values of $f_a$, then axinos are overproduced (blue curve) while for high $f_a\sim 10^{13}$ GeV then axions are over produced (in this case where $\theta_i=1$, purple curve). 
For the case $f_a=10^{11}$ GeV (green curve), then an axino-dominated relic density can be found in accord with the measured DM abundance.
The red curve can also give accord with the measured DM abundance, but this time for the case of dominant axion production (and where $\theta_i$ must be dialed slightly downward for complete accord)..

\section{Conclusions} 
\label{sec:conclude}

In this paper, we have explored possibilities for supersymmetric dark matter in natural SUSY models which are characterized by low 
$\Delta_{EW}\alt 30$ with a PQ solution to the strong CP problem,
and its concomitant axion. We are motivated by the lack of a WIMP signal
at multi-ton-scale noble liquid detectors, such as the recent strong limits from LZ which require a spin-independent WIMP-proton
cross section $\sigma^{SI}(\tchi p) <5\times 10^{-48}$ cm$^2$ for
$m_{\tchi}\sim 200$ GeV (approaching the neutrino floor/fog).
The LZ bound even affects DM models with a depleted WIMP abundance such
as natural SUSY with mixed $a\tchi_1^0$ dark matter where the bulk of 
DM is axions.
An alternative to the usual assumption of a neutralino as LSP is the
possibility of an axino LSP. 
We examine this mainly in the context of the SUSY DFSZ axion model. 
For the bulk of $f_a$ values, the $\tchi_1^0\to \ta Z,\ \ta h$ decay occurs before the onset of BBN.
The $\ta$ can be produced either thermally or non-thermally, and
TP tends to restrict one to very light axino masses $\ll 1$ GeV.
We calculate the relic abundance of mixed $a\ta$ dark matter and found two solution regions: 1. at $f_a\sim 10^{11}$ GeV where axinos comprise the bulk of DM and are likely {\it warm}, and 2. at higher $f_a$ values
$\sim 10^{12-13}$ GeV where TP-axinos are suppressed and where axions 
comprise the bulk of DM. 
This is the more engaging solution since the axions would comprise 
{\it cold} DM. 
While the measured value of the Higgs boson mass favors gravity-mediation (via the expected large trilinear soft breaking terms), 
supergravity calculations tend to favor $m_{\ta}\sim m_{soft}\sim m_{3/2}$, although under special conditions the axino mass can be much lighter. 
A signature for the $\tchi_1^0$ of this scenario would be the detection of delayed neutralino decays in long-lived particle (LLP) search 
experiments\cite{Curtin:2018mvb,Barenboim:2014kka,Jeanty:2026etw}.
The overall DM detection expectations for our natural SUSY model 
with mixed $a\ta$ dark matter are rather similar to that of the 
recent SUSY models with all axion DM, where the WIMPs decay to SM particles via RPV modes\cite{Baer:2025oid,Baer:2025srs}: in both, one only expects axion haloscope detection of SUSY DFSZ axions with a diminished $a\gamma\gamma$ coupling. 
However, the axion-only and mixed $a\ta$ models can be distinguished if
the long-lived NLSP decays can be distinguished at LLP experiments since
in one case the $\tchi_1^0$ decays via RPV-modes while in the case
examined here the decays are to $\ta Z$ and $\ta h$.

\bigskip\bigskip

{\it Acknowledgements:} 
HB and KZ gratefully acknowledge support from the Avenir Foundation.

\bibliography{axino}
\bibliographystyle{elsarticle-num}

\end{document}